# Implications of MBTI in Software Engineering Education

**L. F. Capretz**
University of Western Ontario
Dept. of Electrical & Computer Engineering
London, Ontario, N6G 1H1, Canada
<lcapretz@acm.org>

*"No one has to be good at everything."*
– I.B. Myers

**Abstract**
A number of approaches exist to aid the understanding of individual differences and their effects on teaching and learning. Educators have been using the Myers-Briggs Type Indicator (MBTI) to understand differences in learning styles and to develop teaching methods that cater for the various personality styles. Inspired by the MBTI, we developed a range of practices for effective teaching and learning in a software engineering course. Our aim is to reach every student, but in different ways, by devising various teaching approaches.

## 1. Introduction

The primary goal of teaching is to help students learn. Educators have long believed that it should be possible to use the same instructional methods to teach all students. For many years, research on instruction and teacher behavior was directed to that elusive end. Nowadays, we know that students differ greatly in how they learn. This can create harmony or discord for individual students, depending on whether the student's approach to learning matches the teacher's approach to teaching. Although there are some teaching strategies useful to a whole class, the differences among students make it necessary to diversify those teaching strategies.

Walker [1] states that he knows several computer science teachers who entered this career path, at least in part, because they wanted to act; they wanted an outlet for some form of career involving acting, he explores the idea of promoting learning through elements of theater, like dramatics, stage effects and entertainment. Fortunately, we do not need to go that far for two main reasons: firstly, we can be excellent teachers without acting, and most importantly we will not be reaching all students by acting only.

Many teachers still believe that being fair means treating all students equally. If this translates into using the same approach with every student or treating students identically, then problems are likely to arise for many students who may feel left out because of teacher's choice of classroom activities biased by his or her own teaching style. Once the natural and healthy differences that exist in students are fully understood, teachers can appreciate that being fair really means providing equal opportunities for each student to learn in the manner that best suits his or her own natural learning style.

We base the learning preferences described in this article on the concepts of psychological types developed by the Swiss physician-psychologist Carl Jung. He had the insight that we could identify people by their different - and equally legitimate – preferences that influence the ways in which our minds perceive and organize daily experiences. Myers [2] had the vision to apply that knowledge, determining how people take in information, make decisions, and communicate thoughts and feelings. The Myers-Briggs Type Indicator (MBTI) bases its value on Jung's theory that people with different personality profile will organize information and perceive the world in different ways. The theory of psychological type has the power to transform human relationships, in particular the teacher-student dynamics.

The MBTI is an instrument designed to measure four dimensions of an individual's personality (more on MBTI can be see at www.capt.org). Shortly, MBTI includes four internal scales related to characteristic or preferred ways of becoming aware, reaching conclusions, decision making and general orientation to a private inner world or external world of actions. They call there dimensions introversion (I) and extroversion (E), sensing (S) and intuition (N), thinking (T) and feeling (F), perception (P) and judging (J), respectively. In other words, Es prefer to work interactively with a succession of people, whereas Is prefer work that permit some solitude. Ns prefer working in a succession of new problems and Ss prefer working with detail. Ts want work that requires logical thinking, whereas Fs want work that provides service to people. Js prefer work that imposes a need for order, whereas Ps prefer work that requires adapting to changing situations. We all have personality qualities of each scale or parameter; we simply prefer some qualities or are more comfortable with some styles than others, just as right-handers are more comfortable with the right hand, but sometimes use the left hand.

Summarizing, the MBTI sorts these four sets of preferences, one from each pair, to filter out a person's preferred type. Hence, a person's four preferences indicate which of the 16 personality types he or she fits, as shown in Table 1. Philosophically, this system of classification places an equal value on all 16 types, respects the differences among people, and explains their varying points of view. If the MBTI results show that a person is ISTP, then the terminology is to suggest that the person *prefers* ISTP, not that the person *is* an ISTP. No type is better than any other; the various types are gifts differing.

Table 1: The 16 MBTI types

| ISTJ | ISFJ | INFJ | INTJ |
|------|------|------|------|
| ISTP | ISFP | INFP | INTP |
| ESTP | ESFP | ENFP | ENTP |
| ESTJ | ESFJ | ENFJ | ENTJ |

Understanding learning differences and how they function in the classroom is important to both students and teachers. First teachers must understand their own preferences, how these preferences affect their assumptions about what constitutes effective learning and teaching, and how these assumptions affect their teaching and relationships with students. Second, teachers must be familiar with the learning preferences of their students and with the teaching strategies and learning activities that are most effective in dealing with these preferences. With a greater self-understanding and knowledge of learning preferences, teachers can more successfully design instruction for an entire class, as well as work more effectively with individual students.

## 2. Making Connections in the Classroom

The majority of university faculty members fall further along the scale toward the introvert side than do the majority of university students; research has found that the majority (65%) of faculty members in universities to be intuitives (N), although sensing (S) types dominate applied fields such as engineering and business [3]. Indeed, INTJ and ISTJ are the most common type among university professors. By the way, the majority of elementary and high schoolteachers are ESFJ.

Software engineering attracts significantly more thinking that feeling types. Thinking types in theory are motivated to work with concepts and materials which follow the rules of logic and cause-effect; software engineering students and practicing software engineers have more judging types than perceptive types [4]. We predicted that J students who are goal-oriented and who value systems and order may have an easier time in software engineering programs than P students who value a more adaptive or spontaneous approach.

Schools also have about even numbers of sensing and intuitive types, although engineering schools with high prestige have about two-thirds intuitives [5]. In theory, intuitive types have a greater interest in dealing with material which is abstract and symbolic, whereas the sensing student enjoys details, examples, experiences and well-learned routines The relatively even balance between sensing and intuitive types has important implications for software engineering education because their learning styles are so different. It is not easy to motivate and communicate at the same time to students who prefer hands-on learning presented in a structured way and students who prefer to focus on theory in a global way.

*2.1 Helping Extraverts and Introverts*
Teachers can conduct classes with opportunities to talk and problem solve aloud or in groups. Extraverts often learn better when they can talk aloud about the concepts they have just heard in lecture. They learn best when they have action projects before or accompanying the lecture portion. In on of my lectures, immediately after a lesson on software design, I asked to students to come up with a quick design for a weather system. I divided the students into groups so that in each group, all were extraverts or all were introverts. The groups with extraverts enjoyed the exercise a lot more that the introverts and reached a better design solution in shorter time. I believe that given time and opportunity to the introverts' groups to do the exercise as homework, they would be able to work out good solutions as well.

Suggested Tasks for Extraverts
The task objective is to understand more clearly the difficulties of carrying out the requirements specification for a software system. The students are divided into groups of four people, in which two of them act as users (or clients), while the other two act as systems analyst. A possible scenario for the above role-play exercise is where a multi-screen cinema complex has decided that it is time to replace its current manual ticket issue system with a new state-of-the-art computer system.

Suggested Tasks for Introverts
As they need time, introverts require quiet and space for internal processing after receiving an assignment. Quiet and space allow them some private time to reflect on the assignment and organize their thoughts before expecting participation. A good task could be to make a list of all software development tools that you have used: a) Classify them as stand-alone or integrated tools; b) Which activities of the software life cycle each one of them supports?

*2.2 Challenging Sensing and Intuitive Students*
Sensing students favor understanding from "trying it out" compared with intuitive students who are more inclined to "think it through." However, intuitive teacher find easier to deal with concepts than facts and prefer teaching courses "dealing with ideas and theories" rather than "real life situation." For effective teaching, it is important for faculty to acknowledge their own inclination towards intuition and to make conscious effort to recognize the learning preferences of their sensing students by frequently introducing specific examples, facts, details, and practical applications. Therefore, the sensing students will profit more from a software engineering course that gives them the chance to come up with a real-world design using a particular methodology rather than just listening to the main formalities dictated by a design methodology.

Suggested Exercises for Sensing
As they rely on experience rather than theory, provide sensors with two or three practical examples each time they face a new concept. Use audiovisuals, like movies and models; straight lectures usually are not enough to attract the attention of these students. *Exercise 1*: Comment on the similarities and differences between software design and hardware design.

Suggested Exercises for Intuitives
As they need opportunities to be creative and original, challenge intuitive students with problem-solving activities for which there are multiple solutions or different perspectives. *Exercise 2*: Write down a list of reason in favor of using any standardized design description (e.g. UML), and a list of reasons against standardize the same form of description. *Exercise 3*: When they destroyed the Ariane-5 rocket, the news made headlines in France. The *Liberation* newspaper called it "A 37-billion-franc Fireworks Display" on the front page. What is the responsibility of the press when reporting software-based incidents?

## 2.3 Reaching the Thinking and Feeling Types

Software engineers need not only a broad-based technical competence but also the ability to cope with societal change and personal relationships. They need an appreciation of society's ethical problems and the interpersonal skills to work effectively in groups towards a common solution. Therefore, we need feeling types as software engineers. F students who may find difficult to go through a software engineering course might be retained if teaching is enhanced to encompass their preferred learning styles. Specific addition to courses might include more discussion of design aesthetics, ethics, social, and human factors. We deal with this particular issue in the software engineering course. Two lectures in the course (Human Factor in Software Engineering and Egoless Programming) have been introduced to appeal more to students with the F personality preference.

Suggested Assignments for Thinking
As they excel in inductive reasoning, and perform well when there is a single correct answer, a possible assignment: A well-known word processor consists of a million lines of code. Calculate how many programmers a company would need to write it, assuming that they must complete the project within two years. Given that they are each paid $50,000 per year, what are the costs of that development? (Remember that the average programmer productivity is 20 lines of code per day).

Suggested Assignments for Feeling
As they as skilled in understanding other people, feeling types provide opportunities for friendly interaction, support, and positive feedback. *Assignment*: Suppose you are the manager of a software development project. One of the team members fails to meet the deadline for the coding and testing of a module. What do you do? For the same software project, three months before the software is due to be delivered, the customer requests a change that will require massive efforts. What do you do?

## 2.4 Dealing with Judging and Perceiving Types

Research has shown that the majority of teachers holds preference for judging, and thus demonstrates biases for order and structure in the classroom. A teacher use previous success to reinforce the learner to progress in a systematic manner toward a specific outcome. Teachers can also use a mixed system of instruction consisting of sequentially progressive tasks designed as highly individualized learning activities. Under such a scheme, students determine their own rate and amount of learning, considering their preferences, as they progress through a series of instructional tasks. With this method, the teacher acts as a motivator using cues and feedback on a current activity, so that the student would take up a task, learn it, and move on to the next activity.

Suggested Activities for Judging
As judgers like schedule and predictability, closure of one topic before moving to the next, provide them with a course outline, showing topics covered in each grading period. Use a marking system that recognizes and honors individual achievement. For instance, to pass the course, a student must design and implement a prototype for a small software system. Each student should carry out the design, coding, and testing or the system. They should prepare progress reports during the course and a final report at its completion. Each student must deliver a public lecture on the work performed, followed by a demonstration on the prototype developed. The marking system might be: project proposal (3 weeks-10%), design walkthrough (10 weeks-10%), mid-term design report (2 weeks-20%), implementation (10 weeks-10%), deliver a public lecture (2 weeks-10%), demonstration (after 1 week-10%), and final report (2 weeks-30%).

Suggested Activities for Perceiving
As perceivers perform well when required to quickly adapt to immediate circumstances, allow some flexibility as too many rules weigh heavily on this type of student. Perceiving students could be helped by teaching them to work back-wardly from deadlines, by helping them determine the latest date at which a project can be started and still meet expectations; or even allow some deadline flexibility. Teachers should enforce a marking system that rewards students for maintaining a desirable pace and penalize them for failing to do so. Students' progress improves and learning becomes unhindered when teachers use pacing bonuses or penalties. Such a scheme can be easily applied to project courses; indeed, it has been followed in a course named software engineering design at the University of Western Ontario, and has been demonstrated to be extremely effective in producing significant gains for perceiving student, and increased teacher's freedom.

## 3. Final Remarks

Adjusting instruction to accommodate the learning styles of different types of students can increase both achievement and the enjoyment of learning. The MBTI and its inferences provide a way to conceptualize a student as an organized dynamic personality, which predisposes each student to certain ways of behaving and gives the student a unique learning pattern.

MBTI has proved to be a useful instrument for understanding student learning preferences and has enable comparisons of the learning preferences for various personality types. Regarding learning styles, there is no one best combination of characteristics, since each preference has its own advantages and disadvantages. Therefore, it is a fallacy to think that professors can devise a single teaching technique that would always appeal to all students at the same time.

Software engineering faculty should recognize that their classes contain all types of learners. Hence, effective instruction should try to make some appeal to each learning style for some of the time in a balanced fashion. That means incorporating activities that require reflection and occasional discussion. Challenge them with problem solving exercises involving abstraction and practice; encourage them to see the tree as well as the forest; give them the opportunity to develop a personal (feeling) touch and whenever possible, tolerate deadline flexibility to cater for the needs of the perceiving types. The type theory provides a way of dealing with these issues.

In closing, we remind you that all types choose software engineering. Some types are more likely to stay within the field while others leave. Even so, software engineering is losing some atypical students who tried our wares and then sought more fitting studies; it means that we are losing some students of the types which can be important in transforming software engineering into a more user-oriented field and in finding new directions for software engineering in the future. If we can find ways to value the diversity among students, help them to go through the barrier of type and reach niches in software engineering where they will fit and feel valued, we should thrive to provide alternatives to retain them and enrich our profession.

## References

[1] Walker, H. M. Teaching and a sense of the dramatic. *SIGCSE Bulletin*, 33(4):16-17, December 2001.
[2] Myers, I. B., McCaulley M. H., Quenk N. L. and Hammer A. L. *Manual: A Guide to the Development and Use of the Myers-Briggs Type Indicator*. Consulting Psychologists Press, Palo Alto (CA), 1998.
[3] Provost, J. A. and Adams S. *Applications of the MBTI in Higher Education*. Consulting Psychologists Press, Palo Alto (CA), 1987.
[4] Capretz, L. F. Personality types in software engineering. *International Journal of Human Computer Studies*, submitted February 2002.
[5] Rosati, P. Specific differences and similarities in the leaning preferences of engineering students. *29th ASEE/IEEE Frontiers in Education Conference*, #1544, 1999.

Reviewed Papers

---

# Call for Participation

ACM

# International Student Research Contest

<www1.acm.org/spost/>